\documentclass[epj,showpacs]{svjour}
\usepackage{latexsym}
\usepackage{graphics}
\usepackage{latexsym}
\usepackage{epsfig}
\usepackage{amssymb}
\usepackage{enumerate}
\usepackage{amsmath}
\usepackage{booktabs,dcolumn,caption}

\usepackage{latexsym}
\usepackage{epsfig}
\usepackage{amssymb}
\usepackage{enumerate}
\usepackage{amsmath}
\usepackage{xcolor}
\usepackage{verbatim}

\newcommand{\ba}{\begin{eqnarray}}
\newcommand{\ea}{\end{eqnarray}}
\newcommand{\be}{\begin{equation}}
\newcommand{\ee}{\end{equation}}
\begin{document}
\title{Traversable wormholes with vanishing sound speed in $f(R)$ gravity}

\author{Salvatore Capozziello\inst{1,2,3,4} \thanks{\emph{Present address:} capozziello@na.infn.it} \and Orlando Luongo\inst{5,6,7} \thanks{\emph{Present address:} orlando.luongo@unicam.it} \and Lorenza Mauro\inst{8,9} \thanks{\emph{Present address:} lorenza.mauro@lnf.infn.it}}

\institute{ Dipartimento di Fisica "E. Pancini", Universit\`a di Napoli "Federico II", Via Cinthia, I-80126 Napoli, Italy. \and Scuola Superiore Meridionale, Largo S. Marcellino 10, I-80138, Napoli, Italy \and Istituto Nazionale di Fisica Nucleare (INFN), Sezione di Napoli, Via Cinthia, Napoli, Italy. \and Laboratory of Theoretical Cosmology, Tomsk State University of Control Systems and Radioelectronics (TUSUR), Tomsk, Russia.\and Physics Division, University of Camerino, Via Madonna delle Carceri, Camerino, Italy.\and Dipartimento di Matematica, Universit\`{a} di Pisa, Largo Bruno Pontecorvo, Pisa, Italy. \and National Nanotechnology Laboratory of Open Type (NNLOT), Al-Farabi Kazakh National University, Al-Farabi av. 71, 050040 Almaty, Kazakhstan. \and Dipartimento di Matematica e Fisica, Universit\`{a} degli Studi ``Roma Tre'', Via della Vasca Navale, Roma, Italy \and Istituto Nazionale di Fisica Nucleare, Sezione di Frascati, Via Enrico Fermi, Italy.}
\date{Received: \today / Revised version: \today}

\abstract{
We derive exact  traversable wormhole  solutions in the framework of $f(R)$ gravity with no exotic matter and with stable conditions over the geometric fluid entering the throat. For this purpose, we propose  power-law  $f(R)$ models and two possible approaches for the shape function $b(r)/r$. The first approach makes use of an inverse power law function, namely $b(r)/r\sim r^{-1-\beta}$. The second one adopts Pad\'e approximants, used to characterize the shape function in a model-independent way. We single out the $P(0,1)$ approximant where the fluid perturbations are negligible within the throat, if the sound speed vanishes at $r=r_0$. The former  guarantees an overall stability of the geometrical fluid into the wormhole. Finally we get suitable bounds over the parameters of the model  for  the  above discussed cases. In conclusion, we find that  small deviations from General Relativity give stable solutions.
}

\maketitle

\section{Introduction}\label{intro}

Wormholes are solutions of  field equations in General Relativity (GR) and in several theories of gravity. They can be interpreted as ``short-cuts'' connecting different  space-time regions \cite{visser} and so they represent hypothetical tunnels between two asymptotic regimes of the same space-time \cite{UNO0}. In the most accredited scenario, the wormhole short cut path is traversable through a \emph{minimal surface area called wormhole throat}. The simplest approach  showing these features is the so called  Einstein-Rosen bridge coming from the connection of two Schwarzschild solutions \cite{UNO11,UNO12,UNO13,UNO14}. It is characterized by spherical symmetry and by the presence of an event horizon. This implies that any observer,  trying to cross the wormhole throat,  inevitably falls into the singularity \cite{einsteinrosen}. Hence, the metric itself \emph{a priori} prevents the traversability due to the singularity. Consequently, to heal this issue,  one can consider non-singular metrics defined for every radial coordinates \cite{MorrisThorne1}. However, if the Birkhoff theorem is valid and matter fields are included in a non-vanishing energy-momentum tensor, this approach leads to  severe bounds at the wormhole throat. There,  the condition $\tau_0>\rho_0 c^2$ must hold, i.e. radial tension might be large enough to exceed the total mass-energy density \cite{MorrisThorne1}. Consequently the energy-momentum tensor violates the null energy condition (NEC) at the throat \cite{visser,energy}, $T_{\mu\nu}k^\mu k^\nu<0$, in naive analogy to some cosmological contexts, see e.g. \cite{cosmology,DUE}. Thus, in this \emph{standard approach}, i.e. in the framework of GR, one is forced to take a negative energy density and pressure. This exotic landscape provides a structure that can be \emph{traversable}, albeit it is not clear how matter could exhibit negative energy density and pressure \cite{TRE}. In other words, standard matter cannot be used to achieve stable wormhole solutions in the framework of GR.

In this respect, several approaches have been proposed   to alleviate the problem. They focus mainly in considering  exotic forms of  matter to overcome this strange behavior. Conversely, extended and/or modified theories of gravity are natural suites where this can be addressed retaining standard matter \cite{QUATTRO,REPORT,DARKCURV,CINQUE,SEI}. In fact, in these scenarios, the above conditions  do not apply directly to matter and so, in lieu of imposing exotic conditions over the energy and pressure, one can take geometry to play the role of exotic matter \cite{SETTE}. This mimics the wormhole properties through higher-order curvature terms and/or effective field theories that can be mapped into Lagrangians extending the Hilbert-Einstein one. 

In this paper, in analogy to GR,  we consider a spherically symmetric metric with two asymptotically flat regions. In particular, we take in to account $f(R)$ gravity theories, in metric formalism, and we assume time-independent metric coefficients, using the widely-consolidate Morris-Thorne space-time. We thus determine exact solutions of traversable wormhole, without violating the signs of energy and pressure, postulating a power-law form for $f(R)$, i.e. $f(R)=f_0R^{1+\epsilon}$ where $\epsilon$ is a real number. Immediately, GR can be recovered in the limit $\epsilon\rightarrow0$. In this perspective, we can control  deviations from GR  and the role of geometric terms in stabilizing the wormhole solutions.

In this context, two classes of stable and traversable wormhole can be recovered.

In the first case, the throat is assumed  as an inverse power of the radial coordinate. In the second case, we consider a parameter $\alpha$ controlling the size of the throat.  It is worth noticing  that these solutions can be recovered by assuming simple rational series, made in terms of $(0,1)$ Pad\'e polynomials. We thus provide a physical interpretation over these choices and  investigate the physical properties associated to them. In  this regard, we impose the fluid perturbations passing through the throat are negligibly small. This condition is achieved if  the sound speed is vanishing during the fluid evolution. This feature cannot be found in  GR, albeit it gets suitable constraints over the free coefficients of our wormhole picture. In particular, we show that if the sound speed is zero to guarantee stability, even the Starobinsky scalaron \cite{starob} (with $\sim R^2$) is not fully recovered within this scheme, leading to solutions with $\epsilon\neq 0$. It means that  GR cannot be trivially recovered by construction.  We thus discuss which  power law intervals are allowed under this scheme and  discuss the corresponding physical implications. 

The layout of the paper is the  following. In Sec. II, we take into account $f(R)$ gravity with spherical symmetry  assuming a Morris-Thorne  metric. In the same section, we introduce the basic ingredients of our approach. Thus, we consider  power law $f(R)$ considering deviations with respect to GR.  In particular, we impose the form of the metric functions and  introduce the Pad\'e expansion. In Sec. III,  the stability condition,   given by the vanishing sound speed, is discussed:  we find traversable and stable wormholes and  constrain the parameters $r_0, \beta$ of the exact solutions  and $\epsilon$ of the related gravity model.  Discussion and conclusions are reported  in Sect. IV.


\section{$f(R)$ wormholes with rational shape  function}
\label{fR}

Extended theories of gravity are built up through effective Lagrangians and can present higher-order terms in curvature invariants  that could mimic the exotic behavior of matter within a wormhole \cite{REPORT,fR1}. In principle, they are extensions of Einstein theory where GR is a particular case or it is recovered as soon as higher-order terms reduce to $R$. 
Here we consider a straightforward extension which is  $f(R)$ gravity in the metric formalism \cite{REPORT,fR2}, that is
$S=\int{d^4x\sqrt{-g}[f(R)+\mathcal{L}]}$ 
 where $f(R)$ is a generic function of the curvature scalar $R$, $g$ is the determinant of the metric tensor and $\mathcal{L}$ is the Lagrangian of standard matter minimally coupled to gravity.

Varying with respect to the metric implies the following fourth-order field equations \cite{CURV,simmsferfR}:
\begin{equation}
\label{eqcamp}
G_{\mu\nu}=\frac{1}{f_R(R)}\left\{\frac{1}{2}g_{\mu\nu}\Bigl[f(R)-Rf_R(R)\Bigr]+f_R(R)_{;\mu\nu}-g_{\mu\nu}\Box f_R(R)\right\}+\frac{T_{\mu\nu}^{(m)}}{f_R(R)}\,,
\end{equation}
where $T_{\mu\nu}^{(m)}$ is the stress-energy tensor of ordinary matter.
The right hand side of (\ref{eqcamp}) can be regarded as an effective stress-energy tensor $T_{\mu\nu}^{(eff)}$, given by the sum of $T_{\mu\nu}^{(m)}$ and a \emph{curvature fluid} energy-momentum tensor $T_{\mu\nu}^{(curv)}$, sourcing the effective Einstein equations\footnote{This interpretation is based on the fact that further degrees of freedom coming from higher-order gravity can be recast as an effective perfect fluid which reduces to the standard energy-matter one as soon as  GR is recovered. For a rigorous demonstration, see \cite{Mantica1,Mantica2}.}.

A static and spherically symmetric wormhole solution is 
\begin{equation}
\label{metric}
ds^2=e^{2\Phi(r)}dt^2-\frac{1}{1-b(r)/r}dr^2-r^2d\Omega^2,
\end{equation}
which is the so-called  Morris and Thorne metric \cite{MorrisThorne1}. 
Eq. \eqref{metric}  characterizes a wormhole with the following features: $(i)$ the space-time is  static and  spherically symmetric; $(ii)$ the throat has  a minimal surface connecting two asymptotically flat regions; $(iii)$ there is no Killing horizon and then   two-way travels are enabled. The physical realization of such criteria depends on the gravitational forces,  the  proper time for crossing and astrophysical scales where possible  wormholes are expected \cite{Vittorio}. 

In this picture, $b(r)$ and $\Phi(r)$ are functions of the radial coordinate and they are denoted respectively as \emph{shape} and \emph{redshift functions}. 
The radial coordinate $r$ ranges from a minimum and a positive value $r_0$, defining the  wormhole throat, to infinity.
In order to avoid the presence of event horizons, one imposes  that $\Phi(r)$ is finite at any $r$. It is possible to construct asymptotically flat space-times, where $b(r)/r \rightarrow 0$ and $\Phi \rightarrow 0$ as $r \rightarrow \infty$. A fundamental ingredient in wormhole physics is the the so-called {\it flare-out condition} of the wormhole throat $b(r_0)=r_0$ \cite{MorrisThorne1} given by the condition $(b' r-b)/2b^2 < 0$. In GR, the latter condition implies that through the Einstein field equation, the stress-energy tensor violates the NEC at the throat, i.e., $T_{\mu\nu}k^\mu k^\nu |_{r_0} < 0$.

In  this paper, we consider:
\begin{equation}
\label{redshift}
2\Phi(r)=\frac{r_0}{r}
\end{equation}
that resembles a Newtonian potential for $\Phi(r)$ in analogy to black hole physics  \cite{capozzlobo,rosalobo}. The ratio $\frac{b(r)}{r}$ is debated and its form is \emph{a priori} unknown. Finding out the forms of $\Phi(r)$ and $b(r)$ from the field equations means to derive a Morris-Thorne-like wormhole solution. Here, we follow two physically-motivated strategies imposing:

\begin{subequations}
\begin{align}
\frac{b(r)}{r}&=\left(\frac{r_0}{r}\right)^{1+\beta}\label{shape}\,,\\
\frac{b(r)}{r}&=\frac{r_0}{1+\alpha\,r}\,,\label{shape2}
\end{align}
\end{subequations}
with $\beta\in\mathbb{R}$ and $\beta +1>0$ and $\alpha\equiv\frac{r_0-1}{r_0}$ to guarantee that at $r=r_0$ the wormhole is not singular.  

Therefore our wormhole metric takes two possible forms:
\begin{align}
 \label{wh_metric}
ds^2&=e^{r_0/r}dt^2-\frac{1}{1-\bigl(\frac{r_0}{r}\bigr)^{\beta +1}}dr^2-r^2d\Omega^2\,,\\
\label{wh_metric2}
ds^2&=e^{r_0/r}dt^2-\frac{1}{1-\frac{r_0}{1+\alpha r}}dr^2-r^2d\Omega^2\,.
\end{align}

In both  cases, the  expressions are polynomials  characterizing $b/r$. To enable stability, one can require that within the whole interval of $r$, the ratio $b(r)/r$ does not diverge. An intriguing proposal has been shown in \cite{rosalobo}, where a shape function of the type $b(r)=r_0\left(\frac{r_0}{r}\right)^\beta \exp\left(-\delta\frac{r-r_0}{r_0}\right)$ has been introduced. As $r-r_0\ll 1$, it is possible to make a Taylor expansion that leads to Eq. (\ref{shape}) that turns out to be an extension of the cases  discussed in \cite{capozzlobo,rosalobo}. Hereafter, Eq. \eqref{shape} will be dubbed phenomenological shape function, to stress that is has been argued from heuristic considerations.

At $r=r_0$, to avoid discontinuities in  the Morris-Thorne metric, we can require the domain to be stable even before $r=r_0$. Thus, one can imagine to expand around $r=r_0$ in terms of rational expansions, made by  Pad\'e functions, widely used in  recent literature  \cite{CINQUE,orlando1,orlando2}. The corresponding ratio, constructed by means of Pad\'e polynomials, changes dramatically the form of  solutions or leaves it unaltered. As a prototype of our recipe, we take into account the simplest Pad\'e expansion \cite{orlando3}. To this end, we recall that the Pad\'e technique is built up from the standard Taylor series, being to lower divergences or singular points. Hence, given a function $f(z)=\sum_{i=0}^\infty c_iz^i$, expanded with a given set of coefficients, namely $c_i$,it  is approximated by means of a $(n,m)$  Pad{\'e} approximant by the ratio  \cite{orlando4}:
\begin{equation}
P_{n, m}(z)=\dfrac{\displaystyle{\sum_{\kappa=0}^{n}a_{\kappa} z^{\kappa}}}{1+\displaystyle{\sum_{\sigma=1}^{m}b_\sigma z^{\sigma}}}\,,
\label{eq:def_PadÃ©}
\end{equation}
where the Taylor expansion matches the coefficients of the expansion up to the highest possible order:
\begin{align}\label{am}
&P_{n,m}'(0)=f'(0)\,,\\
&\vdots	\nonumber\\	
&P_{n,m}^{(n+m)}(0)=f^{(n+m)}(0)\,,
\end{align}
with the additional request $P_{n,m}(0)=f(0)$.

\noindent The numerator is thus constructed to have $n+1$ independent coefficients, whereas in the denominator, it is  $m$, for a total of  $n+m+1$ unknown terms. 

For small radii, rational expressions are essentially indistinguishable from the Taylor one, but, at larger radii,  the convergence radius of rational polynomials is determined by the following practical rule \cite{orlando5,orlando6}:
\begin{itemize}
  \item[{\bf I.}] \emph{The most suitable rational approximation order leads to  the function that maximizes the convergence radius.}
  \item[{\bf II.}] \emph{The most suitable rational approximation minimizes the involved free constants.}
\end{itemize}

Consequently a small number of free parameters is essential to enable the rational approximation to be convergent, providing the compromise between arbitrary-order expansions and minimal number of free parameters in the denominator.


The lowest Pad\'e orders are two: (1,0) and (0,1). They turn out to be the simplest approaches to use in the framework of wormholes. We are forced to take the $(0,1)$ order since it  guarantees that all the other assumptions over the stability of $b/r$ are preserved\footnote{The expansion (1,0) corresponds to a first order Taylor expansion and does not work well to guarantee that, for $r\rightarrow\infty$, ${b\over r}\rightarrow0$.  We have to check this property   for all  cases at the wormhole throat, namely $r_0$,  $b(r_0)/r_0=1$. }. In both cases, the asymptotic conditions are automatically satisfied, i.e. $e^{r_0/r} \rightarrow 1$ and $\frac{1}{1-b(r)/r}\rightarrow 1$ as $r \rightarrow \infty$.

\noindent We are now able to get the corresponding energy conditions and to check whether the above consistency  conditions are satisfied.

Let us  start with the field Eqs. (\ref{eqcamp}) which can be rewritten as : 
\begin{equation}
    \label{Tmn}
T_{\mu\nu}^{(m)}=f_R(R) G_{\mu\nu}-\left\{\frac{1}{2}g_{\mu\nu}\Bigl[f(R)-Rf_R(R)\Bigr]+f_R(R)_{;\mu\nu}-g_{\mu\nu}\Box f_R(R)\right\} \,.
\end{equation}
In order to write energy conditions \cite{visser,energy}, we can choose: 
\begin{align}
\label{energytens}
T^{\mu}{}_{\nu}={\rm diag}(\rho,-p_r,-p_t,-p_t)\,,\quad 
f(R)=f_0R^{1+\epsilon}\,,
\end{align}
where $p_t$ is the tangential pressure, $p_r$ the radial pressure and $\rho$ the energy density, whereas $f(R)$ is  a power law, with $f_0$ dimensional constant\footnote{From now on we set this constant equal to 1.}. 
For $\epsilon\ll1$, it can be written in the form  
\begin{equation}\label{sui}
f(R)\sim R+\epsilon R\ln R+{\cal O}(\epsilon^2)\,,
\end{equation}
corresponding to the GR  plus a correction. Clearly this form is useful to control little deviations with respect to the standard Einstein theory. Recently, this approach  revealed particularly useful to study  compact objects, like neutron stars and black holes, where deviations with respect to GR can be useful to fit observations  \cite{Astashenok,Farinelli}.

Here, we  adopt a similar approach to investigate which cases correspond to small departures from Einstein's gravity according  to the values of $\epsilon$. As we will see below, $\epsilon$ is constrained in range of  values  providing  wormhole solutions with vanishing sound speed.  In other words, we can state that:\\ 

\noindent {\it Stable and traversable  wormhole solutions  are possible for small deviations of Einstein's gravity in presence of standard perfect fluid matter. }
\\
Starting from \eqref{Tmn} and \eqref{energytens},  we get the components of the energy-momentum tensor  for the generic metric (\ref{metric}): 
\begin{eqnarray}
\label{ro_caso1}
\rho(r)&=&\frac{1}{2} \left[\frac{r\left(r_0^2+r (r_0-4 r) b'(r)\right)-r_0 (r_0+r) b(r)}{2r^5}\right]^{1+\epsilon}\Biggl\{\epsilon-\frac{4 (1+\epsilon) r^3 b'(r)}{r \left(r (4 r-r_0) b'(r)-r_0^2\right)+r_0 (r_0+r) b(r)}+
\nonumber \\
&&+\frac{2 \epsilon (1+\epsilon) r^2 \left(r b'(r)-b(r)\right) \left(r \left(-r_0^2-4 r_0 r+8 r^2\right) b'(r)-4 r_0^2 r+r^3 (r_0-4 r) b''(r)+r_0 (5 r_0+4 r) b(r)\right)}{\left(r_0 (r_0+r) b(r)-r \left(r_0^2+r (r_0-4 r) b'(r)\right)\right)^2}+
\nonumber \\
&&-\frac{1}{\left(r \left(r_0^2+r (r_0-4 r) b'(r)\right)-r_0 (r_0+r) b(r)\right)^3}\Biggl[2 \epsilon (1+\epsilon) r (r-b(r))\Bigl[r_0^2 (r_0+r) \left(5 r_0^2+44 r_0 r+24 r^2\right) b(r)^2+
\nonumber \\
&&+r_0 r b(r) \Bigl(r^2 (r_0+r) \left(\left(3 r_0^2+6 r_0 r-16 r^2\right) b''(r)-2 r^2 (r_0-4 r) b^{(3)}(r)\right)-3 r_0^2 \left(3 r_0^2+24 r_0 r+16 r^2\right)+
\nonumber \\
&&-\left(r_0^4+26 r_0^3 r+60 r_0^2 r^2-152 r_0 r^3-112 r^4\right) b'(r)\Bigr)+r^2\Bigl(4 \left(r_0^5+6 r_0^4 r+2 (\epsilon-1) r^{13}\right)+
\nonumber \\
&&+r(r_0-4 r) \left(r_0^3+20 r_0^2 r+16 r_0 r^2-16 r^3\right) b'(r)^2+r_0^2 r^2 \left(\left(-3 r_0^2-6 r_0 r+16 r^2\right) b''(r)+2 r^2 (r_0-4 r) b^{(3)}(r)\right)+
\nonumber \\
&&+b'(r) \bigl(r^3 (r_0-4 r) \left(\left(-3 r_0^2-6 r_0 r+16 r^2\right) b''(r)+2 r^2 (r_0-4 r) b^{(3)}(r)\right)+
\nonumber \\
&&+r_0^2 \left(r_0^3+24 r_0^2 r+24 r_0 r^2-112r^3\right)\bigr)\Bigr)\Bigr]\Biggr]\Biggr\}\,,
\end{eqnarray}
\begin{eqnarray}
\label{pr_caso1}
p_r(r)&=&\frac{1}{2} \left[\frac{r \left(r_0^2+r (r_0-4 r) b'(r)\right)-r_0 (r_0+r) b(r)}{2r^5}\right]^{1+\epsilon}\Biggl\{\Biggl[-\epsilon+\frac{4 (1+\epsilon) r ((r-r_0) b(r)+r_0 r)}{r \left(r (4 r-r_0) b'(r)-r_0^2\right)+r_0 (r_0+r) b(r)}+ 
\nonumber \\
&&+\left(2 \epsilon (1+\epsilon) r (r_0-4 r) (r-b(r)) \left(r \left(r_0^2+4 r_0 r-8 r^2\right) b'(r)+4 r_0^2 r+r^3 (4 r-r_0) b''(r)-r_0 (5 r_0+4 r) b(r)\right)\right)\Biggr]\times
\nonumber \\
&&\times\Biggl[\Bigl(r_0 (r_0+r) b(r)-r \left(r_0^2+r (r_0-4 r) b'(r)\right)\Bigr)^2\Biggr]^{-1}\Biggr\}\,,
\end{eqnarray}
\begin{eqnarray}
\label{pt_caso1}
p_t(r)&=&\frac{1}{2} \left[\frac{r \left(r_0^2+r (r_0-4 r) b'(r)\right)-r_0 (r_0+r) b(r)}{2r^5}\right]^{1+\epsilon}\Biggl\{\frac{(1+\epsilon) \left(\left(r_0^2+r_0 r+2 r^2\right) b(r)+r \left(r (r_0-2 r) b'(r)-r_0 (r_0+2 r)\right)\right)}{r \left(r_0^2+r (r_0-4 r) b'(r)\right)-r_0 (r_0+r) b(r)}+
\nonumber \\
&&-\epsilon+\frac{2 \epsilon (1+\epsilon) r^2 \left(b(r)-r b'(r)\right) \left(r \left(-r_0^2-4 r_0 r+8 r^2\right) b'(r)-4 r_0^2 r+r^3 (r_0-4 r) b''(r)+r_0 (5 r_0+4 r) b(r)\right)}{\left(r_0 (r_0+r) b(r)-r \left(r_0^2+r (r_0-4 r) b'(r)\right)\right)^2}+
\nonumber \\
&&+\frac{1}{\left(r \left(r_0^2+r (r_0-4 r) b'(r)\right)-r_0 (r_0+r) b(r)\right)^3}\Biggl[2 \epsilon (1+\epsilon) r (r-b(r))\Bigl[r_0^2 (r_0+r) \left(5 r_0^2+54 r_0 r+32 r^2\right) b(r)^2+
\nonumber \\
&&+r_0 r b(r)\Bigl(-r_0^2 \left(9 r_0^2+90 r_0 r+64 r^2\right)-\left(r_0^4+28 r_0^3 r+80 r_0^2 r^2-192 r_0 r^3-160 r^4\right) b'(r)   +
\nonumber \\
&&+r^2 (r_0+r) \left(\left(3 r_0^2+8 r_0 r-24 r^2\right) b''(r)-2 r^2 (r_0-4 r) b^{(3)}(r)\right)\Bigr)+r^2\Bigl(4 \left(r_0^5+8 r_0^4 r+2 (\epsilon-1) r^{13}\right)+
\nonumber \\
&&+r (r_0-4 r) (r_0+2 r) \left(r_0^2+20 r_0 r-16 r^2\right) b'(r)^2+r_0^2 r^2 \left(\left(-3 r_0^2-8 r_0 r+24 r^2\right) b''(r)+2 r^2 (r_0-4 r) b^{(3)}(r)\right)+
\nonumber \\
&&b'(r) \Bigl(r^3 (r_0-4 r) \left(\left(-3 r_0^2-8 r_0 r+24 r^2\right) b''(r)+2 r^2 (r_0-4 r) b^{(3)}(r)\right)+
\nonumber \\
&&+r_0^2 \left(r_0^3+26 r_0^2 r+40 r_0 r^2-160 r^3\right)\Bigr)\Bigr)\Bigr]\Biggr]\Biggr\}\,.
\end{eqnarray}
Here, the form of $b(r)$ is not specified. If we consider the energy-momentum tensor for perfect fluids written in the form 
$T^{\mu}{}_{\nu}={\rm diag}(\rho,-p,-p,-p)
$,  the average pressure is $
    p(r)=\frac{1}{3}\left[p_r(r)+2p_t(r)\right]$, 
and then:
\begin{eqnarray}
\label{pressure2}
p(r)&=&\frac{1}{6} \left[\frac{r \left(r_0^2+r (r_0-4 r) b'(r)\right)-r_0 (r_0+r) b(r)}{2r^5}\right]^{1+\epsilon}\Biggl\{\frac{4 (1+\epsilon) r ((r-r_0) b(r)+r_0 r)}{r \left(r (4 r-r_0) b'(r)-r_0^2\right)+r_0 (r_0+r) b(r)}+
\nonumber \\
&&-\epsilon+\frac{2 \epsilon (1+\epsilon) r (r_0-4 r) (r-b(r)) \left(r \left(r_0^2+4 r_0 r-8 r^2\right) b'(r)+4 r_0^2 r+r^3 (4 r-r_0) b''(r)-r_0 (5 r_0+4 r) b(r)\right)}{\left(r_0 (r_0+r) b(r)-r \left(r_0^2+r (r_0-4 r) b'(r)\right)\right)^2}+
\nonumber \\
&&+2\Biggl[-\epsilon+\frac{(1+\epsilon) \left(\left(r_0^2+r_0 r+2 r^2\right) b(r)+r \left(r (r_0-2 r) b'(r)-r_0 (r_0+2 r)\right)\right)}{r \left(r_0^2+r (r_0-4 r) b'(r)\right)-r_0 (r_0+r) b(r)}+
\nonumber \\
&&+\frac{2 \epsilon (1+\epsilon) r^2 \left(b(r)-r b'(r)\right) \left(r \left(-r_0^2-4 r_0 r+8 r^2\right) b'(r)-4 r_0^2 r+r^3 (r_0-4 r) b''(r)+r_0 (5 r_0+4 r) b(r)\right)}{\left(r_0 (r_0+r) b(r)-r \left(r_0^2+r (r_0-4 r) b'(r)\right)\right)^2}+
\nonumber \\
&&+\frac{1}{\left(r \left(r_0^2+r (r_0-4 r) b'(r)\right)-r_0 (r_0+r) b(r)\right)^3}\Bigl[2 \epsilon (1+\epsilon) r (r-b(r))\Bigl(r_0^2 (r_0+r) \left(5 r_0^2+54 r_0 r+32 r^2\right) b(r)^2+
\nonumber \\
&&+r_0 r b(r) \bigl(r^2 (r_0+r) \left(\left(3 r_0^2+8 r_0 r-24 r^2\right) b''(r)-2 r^2 (r_0-4 r) b^{(3)}(r)\right)-r_0^2 \left(9 r_0^2+90 r_0 r+64 r^2\right)+
\nonumber \\
&&-\left(r_0^4+28 r_0^3 r+80 r_0^2 r^2-192 r_0 r^3-160 r^4\right) b'(r)\bigr)+r^2\bigl(4 \left(r_0^5+8 r_0^4 r+2 (\epsilon-1) r^{13}\right)+
\nonumber \\
&&+r(r_0-4 r) (r_0+2 r) \left(r_0^2+20 r_0 r-16 r^2\right) b'(r)^2+r_0^2 r^2 \left(\left(-3 r_0^2-8 r_0 r+24 r^2\right) b''(r)+2 r^2 (r_0-4 r) b^{(3)}(r)\right)+
\nonumber \\
&&+b'(r) \bigl(r^3 (r_0-4 r) \left(\left(-3 r_0^2-8 r_0 r+24 r^2\right) b''(r)+2 r^2 (r_0-4 r) b^{(3)}(r)\right)+
\nonumber \\
&&+r_0^2 \left(r_0^3+26 r_0^2 r+40 r_0 r^2-160 r^3\right)\bigr)\bigr)\Bigr)\Bigr]\Biggr]\Biggr\}\,.
\end{eqnarray}
In this way, the null energy condition \cite{visser,energy} at the throat, for the metric (\ref{wh_metric}), that is for $\frac{b(r)}{r}=\left(\frac{r_0}{r}\right)^{\beta+1}$, is expressed as:
\begin{equation}
\label{nec}
\rho+p\vert_{r_0}= \frac{(1+\epsilon)\left(\frac{3\beta-1}{2r_0^2}\right)^{1+\epsilon}\left\{6+\beta-3\beta^2(8+\beta)+(1+\epsilon)\left[-5+\beta(1+3\beta(3+\beta))\right]\right\}}{3(1-3\beta)^2}\geq 0\,,
 \end{equation}
while, for the metric (\ref{wh_metric2}), i.e. for $\frac{b(r)}{r}=\frac{r_0}{1+\alpha r}$, it is: 
\begin{equation}
\label{nec2}
    \rho+p\vert_{r_0}=-\frac{ (1+\epsilon) \left(-\frac{r_0+3}{2r_0^3}\right)^{1+\epsilon} \left\{(r_0-1) \left[r_0 (5 r_0+9)-6\right] (1+\epsilon)-2 \left[r_0 \left(3 r_0^2+r_0-15\right)+3\right]\right\}}{3 r_0 (r_0+3)^2}\geq 0\,. 
\end{equation} 
Another aspect to check is  the flare-out condition \cite{flaringout}:
\begin{equation}
    \frac{b'(r)r-b(r)}{2b(r)^2}\Biggl|_{r_0}<0\,,
\end{equation}
which becomes, for the metric (\ref{wh_metric}) at the throat:
\begin{equation}
    \label{flare}
    -\frac{1+\beta}{2r_0}<0\quad \rightarrow\quad \beta>-1\,.
\end{equation}
In the other case, it  is fulfilled for:
\begin{equation}
    \label{flare2}
   -\frac{r_0-1}{2 r_0^2}<0\quad \rightarrow\quad r_0>1 \,,
\end{equation}
provided that $r_0>0$. The consistency of our model is guaranteed in both  cases. These results allow to give necessary conditions on  the function  $b(r)/r$ but they are not sufficient to show that the  form of $b(r)$ is of the form of a polynomial. To ensure this hypothesis, we assume that the sound speed, i.e. the variation of the pressure with respect to the density, is negligibly small \cite{ulla}. Combining this additional requirement, we stabilize the solution as we shall show below.

\section{The stability condition and the sound speed}
\label{sound}

We require the solutions to be stable, besides being  traversable. This reflects to the stability of fluids inside the throat, with the hypothesis of satisfying the above energy conditions.

Thus, let us  consider the perturbation condition by means of  the adiabatic  sound speed, $c_s$, i.e. we assume the sound speed definition in adiabatic perturbations, in analogy to  what happens in fluid dynamics \cite{suono1,suono2}. So, defining the adiabatic sound speed by \cite{suono3,suono4,suono5}:
\begin{equation}
\label{eq:sample19b}
    c_s^2=\left(\frac{\partial p}{\partial \rho}\right)_S,
\end{equation}
we can guarantee how perturbations affect solutions analyzing its value within the throat. Hence, the sound speed is essential to guarantee the viability of  our approximated versions of $b/r$. The above expression for $c_s$ can be specified as 
\begin{equation}
    \label{stability}
    \frac{dp}{d\rho}\Biggl|_{r_0}=0\;.
\end{equation}
Plugging Eqs. (\ref{ro_caso1}) and (\ref{pressure2}) in Eq. \eqref{stability}  and considering ${\displaystyle \frac{b(r)}{r}=\left(\frac{r_0}{r}\right)^{\beta+1}}$, we get the following stability condition for the metric (\ref{wh_metric}):
\begin{eqnarray}
\label{stability2}
&&147-32r_0^8\epsilon(\epsilon-1)(1+\beta)-4(1+\epsilon)^2(1+\beta)[5-3\beta(2+\beta)]^2+\beta\{-428+\beta[-926+3\beta(148+\beta(161+24\beta))]\}+
\nonumber\\
&&-(1+\epsilon)\{45+\beta[-288+\beta(-658+3\beta(60+\beta(95+12\beta)))]\}\Bigl\{3\Bigl[16r_0^8\epsilon(\epsilon-1)(1+\beta)+2(1+\epsilon)^2(1+\beta)[5-3\beta(2+\beta)]^2+
\nonumber \\
&&-55+(1+\epsilon)(1+\beta)[5+\beta(-113+3\beta(-37+\beta(19+6\beta)))]+\beta[166-\beta(-400+3\beta(38+\beta(67+12\beta)))]\Bigr]\Bigr\}^{-1}=0\,.
\end{eqnarray}
Similarly, by considering ${\displaystyle \frac{b(r)}{r}=\frac{r_0}{1+\alpha r}}$, we get the stability condition for the metric (\ref{wh_metric2}):
\begin{eqnarray}
\label{stability3}
&&\frac{1}{3}\Biggl\{-2-r_0 (r_0+3) \Bigl[r_0 \left(r_0^2 (35 (1+\epsilon)-37)-12 r_0 (\epsilon+2)-177 (1+\epsilon)+135\right)+90 (1+\epsilon)-54\Bigr]\Bigl[r_0+
\nonumber \\
&&+36 (7-5 (1+\epsilon))+\left(-714+ r_0\left(393+r_0 \left(389+r_0 \left(-137+r_0 \left(32 (r_0-1) r_0^7-55\right)\right)\right)\right)\right)+
\nonumber \\
&&+r_0(1+\epsilon) \left(354+ r_0 \left(-39+r_0 \left(-163+r_0 \left(-48 r_0^9+48 r_0^8+5 r_0+23\right)\right)\right)\right)+
\nonumber \\
&&2 (r_0-1)(1+\epsilon)^2 \left(36+r_0 \left(-108+r_0 \left(21+r_0 \left(8 r_0^9+25 r_0+90\right)\right)\right)\right)\Bigr]^{-1}\Biggr\}=0\,.
\end{eqnarray}

\begin{figure*}
\centering
\includegraphics[width=0.4\hsize,clip]{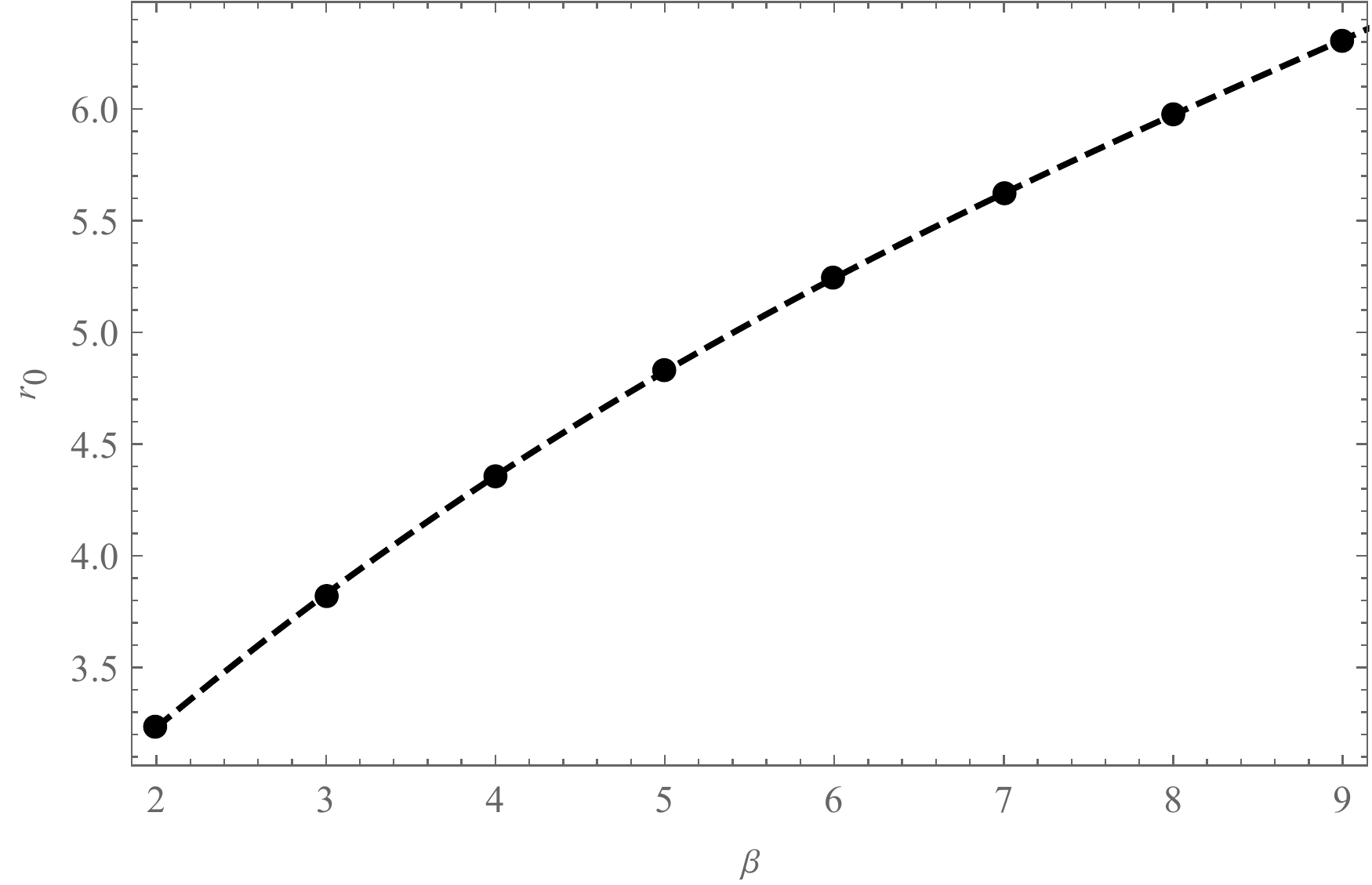}
\includegraphics[width=0.4\hsize,clip]{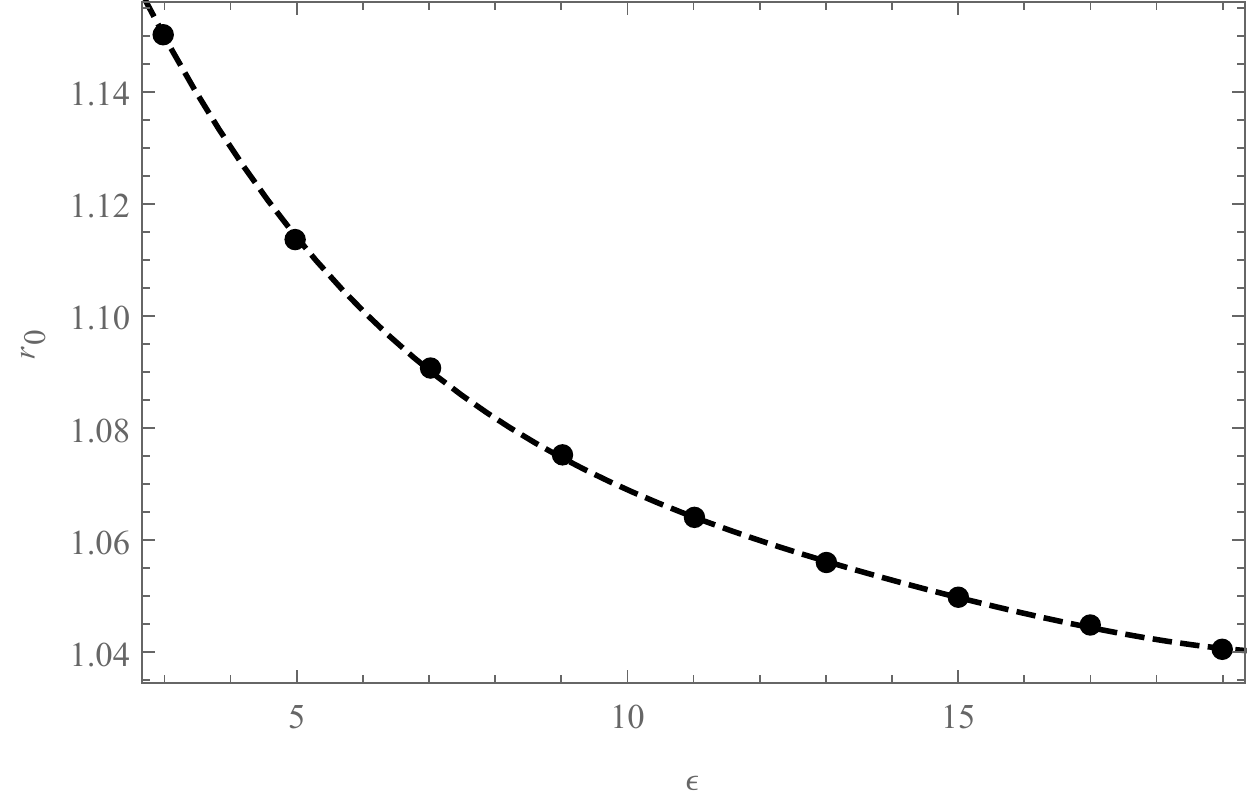}
\caption{Interpolations of  $r_0$ vs $\beta$ and of  $r_0$ vs $\epsilon$. The  curves have been obtained through  {\it Wolfram Mathematica} and represent an extrapolation based on the fitting functions $3.03\cdot 10^{-5} x^4+1.13\cdot 10^{-3} x^3-4.65\cdot 10^{-2} x^2+8.01\cdot 10^{-1} x+1.79 $ and $2.43\cdot 10^{-6} x^4-1.43\cdot 10^{-4} x^3+3.30\cdot 10^{-3} x^2-3.80\cdot 10^{-2} x+1.24$ respectively for the left and right plots.}
\label{fig1}
\end{figure*}

\begin{figure*}
\centering
\includegraphics[width=0.6\hsize,clip]{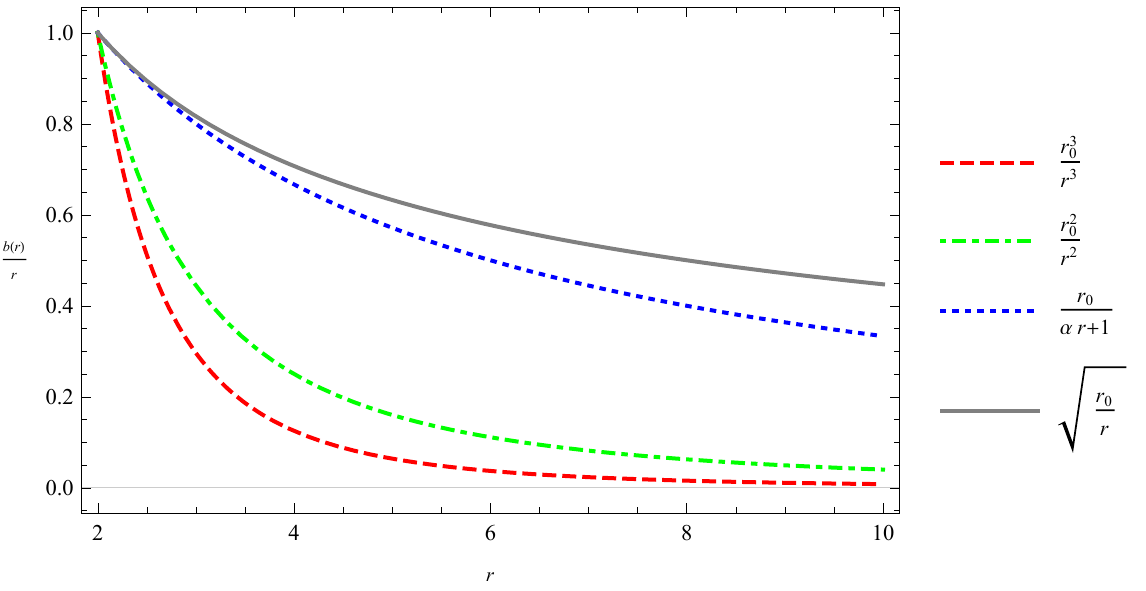}
\caption{ Comparison between the shape functions of our models (red and blue lines),  the \emph{standard  approach} proposed in \cite{MorrisThorne1} (grey line) and the model proposed in \cite{rosalobo} (green line). The value of the $\beta$ parameter chosen for our model (\ref{shape}) is $\beta=2$. Our Pad\'e expansion better adapts to the standard approach than other phenomenological ansatz. }
\label{fig2}
\end{figure*}
We can therefore analyze the consequence of such conditions for our wormhole solutions as reported in the next subsection. The corresponding results are clearly numerical since no analytical solutions can be obtained integrating the above stability conditions\footnote{The assumption $c_s=0$ is also used in astrophysics to guarantee stability of virialized structures. See e.g. \cite{orl}.} coming from $c_s=0$.

\subsection{The wormhole solutions}
\noindent
In the case of metric (\ref{wh_metric}), we have determined a class of wormhole solutions  satisfying the three above conditions, namely the null energy condition (\ref{nec}), the flare-out condition (\ref{flare}) and the stability condition (\ref{stability2}). In particular, once the value of the $\beta$ parameter is fixed, we determine the wormhole throat (as a function of $\epsilon$) and the values of $\epsilon$ that satisfy  Eqs. (\ref{nec}), (\ref{flare}), (\ref{stability2})  and the condition $r_0> 0$. We have summarized the results obtained in Table \ref{tab:1}.
\begin{table}
\centering
\caption{Wormhole solutions for different values of the $\beta$ parameter. Here we adopted the parametrization given in Eq. \eqref{shape}. The allowed range values of $\epsilon$ are also reported.}
\label{tab:1}       
\begin{tabular}{lll}
\hline\noalign{\smallskip}
$\beta$ & $r_0$ & $\epsilon$  \\
\noalign{\smallskip}\hline\noalign{\smallskip}
0          & $\frac{1275}{128}$   & $\frac{1}{5}$       \\
1         & no solution     & no solution           \\
2           & $\sqrt[8]{\frac{5332 \epsilon ^2+14653 \epsilon +150}{96 \epsilon -96 \epsilon ^2}} $     & $-2.986\lesssim \epsilon<-1\cup 0.965\lesssim \epsilon<1$  \\
3         &  $\sqrt[8]{\frac{2 \left(100 \epsilon ^2+317 \epsilon -14\right)}{(1-\epsilon) \epsilon }} $      & $-3.213\lesssim \epsilon<-1\cup 0.800\lesssim \epsilon<1$           \\
4         &  $\sqrt[8]{\frac{89780 \epsilon ^2+289269 \epsilon -9922}{160 (1-\epsilon) \epsilon }} $      & $-3.256\lesssim \epsilon<-1\cup 0.689\lesssim \epsilon<1$           \\
5         &  $\sqrt[8]{\frac{15000 \epsilon ^2+48455 \epsilon -1372}{12 (1-\epsilon) \epsilon }} $      & $-3.258\lesssim \epsilon<-1\cup 0.607\lesssim \epsilon<1$   \\
6         &  $\sqrt[8]{\frac{540988 \epsilon ^2+1744781 \epsilon -42194}{224 (1-\epsilon) \epsilon }} $      & $-3.249\lesssim \epsilon<-1\cup 0.542\lesssim \epsilon<1$           \\
7         &  $\sqrt[8]{\frac{8464 \epsilon ^2+27216 \epsilon -575}{2 (1-\epsilon) \epsilon }} $      & $-3.236\lesssim \epsilon<-1\cup 0.489\lesssim \epsilon<1$           \\
8         &  $\sqrt[8]{\frac{1988100 \epsilon ^2+6370997 \epsilon -119554}{288 (1-\epsilon) \epsilon }} $      & $-3.223\lesssim \epsilon<-1\cup 0.446\lesssim \epsilon<1$           \\
9         &  $\sqrt[8]{\frac{213160 \epsilon ^2+680759 \epsilon -11492}{20 (1-\epsilon) \epsilon }} $      & $-3.210\lesssim \epsilon<-1\cup 0.409\lesssim \epsilon<1$           \\
\noalign{\smallskip}\hline
\end{tabular}
\end{table}
In particular, we note that $\epsilon=\frac{1}{2}$, i.e.  $f(R)=R^{3/2}$,  is obtained in three cases\footnote{$f(R)$ gravity with $1+\epsilon=3/2$ is particularly relevant for cosmological and astrophysical applications. It is related to invertible conformal transformation \cite{CURV}. It allows a curvature interpretation of dark matter phenomena according to MOND \cite{Borka} and the transition from decelerated to accelerated regimes in cosmology \cite{Claudio}. }:
\begin{enumerate}
    \item $\beta=7\Rightarrow r_0\simeq 3.632$;
    \item $\beta=8\Rightarrow r_0\simeq 3.862$;
    \item $\beta=9\Rightarrow r_0\simeq 4.078$.
\end{enumerate}
These correspond to three wormhole metrics, respectively:
\begin{enumerate}
    \item $ds^2=e^{r_0/r}dt^2-\frac{1}{1-\bigl(\frac{r_0}{r}\bigr)^{8}}dr^2-r^2d\Omega^2\;; $
    \item $ds^2=e^{r_0/r}dt^2-\frac{1}{1-\bigl(\frac{r_0}{r}\bigr)^{9}}dr^2-r^2d\Omega^2\;; $
    \item $ds^2=e^{r_0/r}dt^2-\frac{1}{1-\bigl(\frac{r_0}{r}\bigr)^{10}}dr^2-r^2d\Omega^2\;. $
\end{enumerate}
In the case of metric (\ref{wh_metric2}), the validity  of  NEC (\ref{nec2}) points out that $1+\epsilon$ must be an integer. For $\epsilon$ integer and odd, we found no solutions  satisfying also the flare-out condition (\ref{flare2}). Consequently the only solutions that satisfy all three conditions (\ref{nec2}), (\ref{flare2}) and (\ref{stability3}) are those with $\epsilon$ integer and even. Therefore, once the value of $\epsilon$ is fixed, we  determine the corresponding wormhole throat. The results are summarized in Table \ref{tab:2}. We note that  $\epsilon = 0$, implying GR, leads to no solutions as the NEC is not satisfied. The theoretical consequences of our approach are summarized below.
\begin{table}
\centering
\caption{Wormhole solutions for different values of $\epsilon$. Here we adopted the parametrization given in Eq. \eqref{shape2}.}
\label{tab:2}       
\begin{tabular}{ll}
\hline\noalign{\smallskip}
$\epsilon$ & $r_0$   \\
\noalign{\smallskip}\hline\noalign{\smallskip}
0          & no solution      \\
2         & $1.150$             \\
4           & $1.114$     \\
6         &  $1.090$          \\
8         &  $1.075$      \\
10         &  $1.064$      \\
12         &  $1.056$        \\
14         &  $1.050$  \\
16         &  $1.045$ \\
18         &  $1.040$           \\
\noalign{\smallskip}\hline
\end{tabular}
\end{table}

\subsection{Theoretical considerations}

We assumed a power-law form for  $f(R)$, where GR is recovered for $\epsilon=0$. From our analyses, it is possible to provide  two classes of the shape function  $b(r)/r$. The first possibility,  already  adopted in  literature,  represents a class of inverse powers with respect to $r$ \cite{capozzlobo,rosalobo}. This approach departs significantly from the one provided in the original work by Morris and Thorne \cite{MorrisThorne1} as it appears evident from Fig. \ref{fig1}. Even though appealing, these possibilities are therefore disfavored than the Pad\'e expansion that we proposed above. The $(0,1)$ Pad\'e polynomial resembles much more the Morris-Thorne shape function and candidates as a suitable approach that turns out to be \emph{model-independent} in reconstructing $b/r$. The expansion of $b/r$ is constructed by means of rational series. The only dependence from the model occurs as one chooses the order $(n,m)$. This approach is significantly better than \emph{ad hoc} functions postulated at the beginning over $b$. In this respect, it is possible to provide two cases summarized in Tabs. I and II. In Tab. I, we consider the case of  inverse power law approximation. It appears evident that the case $f\sim R^2$ is not recovered  indicating small departures from the Starobinsky scalaron model \cite{starob}.  The inverse solution, i.e.  $1+\epsilon$  negative, is not excluded. In this case the repulsive effects are stronger than the case of positive $1+\epsilon$. On the other hand, the Pad\'e approximation excludes GR as well as the previous case but shows very small departures from $r_0$, indicating moreover that the Starobinsky scalaron is excluded again. In particular, the energy conditions are not fulfilled in the case of odd $\epsilon$. The results are well-suited in the Pad\'e scenario and candidate to reconstruct the shape function without imposing any \emph{ad hoc} functions. In all the aforementioned cases, it is possible to notice that $\epsilon$ is quite small, confirming that only small deviations from GR are permitted as soon as  one considers wormholes in extended theories of gravity with vanishing sound speed.

\section{Final outlooks and perspectives}\label{sect5}
\noindent
In GR,   wormhole solutions  are possible only if exotic matter is considered. In other words, standard matter prevents the wormhole stability and, consequently, its formation. Hence, natural landscapes in which wormholes may exist could be represented by extended and/or modified theories of gravity. There, the wormhole structure can be traversable without considering exotic matter contributions, i.e. a fluid with a negative energy and pressure density violating the energy conditions. 

Here we considered  $f(R)$ gravity. Postulating a  power-law form $f(R)=f_0 R^{1+\epsilon}$, we investigated two possible approaches to characterize the shape function and, in particular, the ratio $b(r)/r$. The first attempt is a phenomenological inverse power law, recovered from widely-investigated approaches in the literature. The second  considers the numerical pathology of the ratio $b(r)/r$ within the throat. We thus introduced the Pad\'e approximant to characterize the shape function in a model-independent way. Our strategy to decide the Pad\'e orders is straightforward: we singled out the simplest approximant that resembles a first order Taylor expansion. To do so, there are two possibilities, i.e. the (1,0) and (0,1) expansions. The first coincides with pure Taylor expansion and is unable to guarantee that $b\rightarrow0$ as $r\rightarrow\infty$. The second possibility, namely  (0,1), fulfills our requirement. Thus, we worked out the wormhole solution under this ansatz, adding the additional requirement of stable  fluids, whose perturbations are negligible inside the throat. For this purpose, we assumed the sound speed to vanish in analogy to cosmological contexts where the sound speed is associated to the fluid evolution. In particular, the sound speed has to vanish on the wormhole throat. Finally, it is possible to constrain the set of coefficients ($\beta$, $\epsilon$,  $r_0$). It is worth noticing that in all the analyzed cases, as soon as GR is recovered, exotic matter is  needed to satisfy wormhole stability and traversability criteria.

In future works, we will extend such a scenario considering a general  Pad\'e approach to characterize the shape function. Other extended/modified theories of gravity will be taken into account besides $f(R)$ gravity.

\section{Acknowledgments}
This work was partially supported by the Ministry of Education and Science of the Republic of Kazakhstan,  Grant: IRN AP08052311. SC and LM acknowledge the partial support of Istituto Nazionale di Fisica Nucleare (INFN) {\it iniziative specifiche} MOONLIGHT2 and QGSKY.

\end{document}